\documentclass[letterpaper,twocolumn,10pt]{article}

\usepackage{ol}
\usepackage{amsmath}
\usepackage{graphicx}

\usepackage{units}
\usepackage{nicefrac}

\newcommand{\nm}[1]{\unit[#1]{nm}}
\newcommand{\um}[1]{\unit[#1]{\ensuremath{\mu}m}}
\newcommand{\V}[1]{{\bf #1}}

\begin{document}

\twocolumn[

\title{Plasmon Assisted Transparency in Metal--Dielectric Microspheres}

\author{Charles Rohde, Keisuke Hasegawa, Miriam Deutsch}

\address{Department of Physics, University of Oregon, 1371 E. 13th St, Eugene, Oregon 97403}

\begin{abstract}
We present a theoretical analysis of light scattering from a layered metal-dielectric microsphere. The system consists of two spherical resonators, coupled through concentric embedding. Solving for the modes of this system we find that near an avoided crossing the scattering cross section is dramatically suppressed, exhibiting a tunable optical transparency. Similar to electromagnetically induced transparency, this phenomenon is associated with a large group delay, which in our system is manifest as flat azimuthal dispersion.  

\end{abstract} 

] %

\noindent 
Nanostructured metallodielectric materials have been extensively studied in recent years due to their promise for new photonic device applications. This has led to the emergence of the new field of \emph{nano-plasmonics}, which addresses the excitation and manipulation of surface plasmon polaritons (SPPs) in these systems~\cite{Barnes2003Surface-plasmon}. In particular, metallodielectrics patterned periodically in three dimensions (3D) exhibit novel dispersion characteristics which rely on coherent scattering of SPPs ~\cite{Moroz,Soukoulis,Baumberg}, potentially allowing sub-wavelength manipulation of light signals. 

An important geometry extensively addressed is that of the 3D spherical plasmonic resonator. The latter consists of a metal nanoshell surrounding a nanoscale dielectric core, and may be designed to allow tuning of the SPP field distributions as well as its resonance~\cite{Halas1}. The large plasmon fields in these systems have proven useful for surface-enhanced Raman spectroscopy~\cite{Halas2}. It has been recently shown that concentric metal-dielectric shells surrounding a meso-scale \emph{metal} sphere (MDM resonators) allow as much as an order of magnitude additional enhancement of the absorption cross section, while preserving the broad tunability of the composite resonance~\cite{Hasegawa2006Enhanced-surface}.

In this Letter we analyze the greatly reduced scattering cross sections of meso-scale MDM particles. The use of plasmonic coatings to reduce dipole scattering from small spheres has been investigated.\cite{Alu} Here we use two-band coupling formalism to show that electromagnetic (EM) transparencies may be achieved in spheres of \emph{any size}. Our system consists of a micron-size metal sphere of permittivity $\epsilon_m$ and radius $R$ surrounded by one concentric sequence of dielectric (permittivity $\epsilon_d$) and metal ($\epsilon_m$) shells of thicknesses $L$ and $T$, respectively. The composite particle is embedded in an isotropic and homogeneous dielectric host with permittivity $\epsilon_0$. We show that with proper design of the metal-dielectric shells it is possible to engineer the dispersion characteristics of a MDM resonator such that the forward scattering cross section is strongly suppressed, resulting in a tunable transparency. By solving the eigenvalue problem we show that the observed transparency is associated with an avoided crossing of the dominant plasmonic bands.

The eigenmodes of a MDM sphere of total radius $W=R+L+T$ are generally obtained by solving a $6\times6$ determinant equation. Although initially cumbersome, once simplified this is rewritten as the $2\times2$ determinant 

\begin{equation}
            \left| \begin{array}{cc} 
                                \left|\V{u}\right|  &     \left|\V{A}\right| \\ 
                                \left|\V{B}\right|  &     \left|\V{V}\right|
                        \end{array}
                    \right| =0
\label{eqn:mdm}
\end{equation}

\noindent The solutions to $\left|\V{u}\right| = 0$, where

\begin{equation}
\V{u} \equiv \left[ \begin{array}{cc} 
                          \eta_m \xi_\ell'(k_0 W)   & \psi_\ell'(k_m W) \\ 
                               h_\ell(k_0 W)            &     j_\ell(k_m W)
                       \end{array}
                 \right]
    \label{eqn:ss}            
\end{equation}

\noindent give the eigenfrequencies of a metal sphere of radius $W$ embedded in a dielectric with permittivity $\epsilon_0$.\cite{Bohren1983Absorption} Similarly, the solutions to $\left|\V{V}\right| = 0$, where \V{V} is given by

\begin{equation}
\setlength\arraycolsep{4pt}
\left[
\begin{array}{@{}cllc@{}}
   \xi_\ell'(k_m S)  &   \psi_\ell'(k_d S) & \xi_\ell'(k_d S) &   0\\[5pt]
        h_\ell(k_m S)        &   j_\ell(k_d S) \eta_d    & h_\ell(k_d S) \eta_d   &    0\\[5pt]
        0                    &  \psi_\ell'(k_d R)   & \xi_\ell'(k_d R)          & \psi_\ell'(k_m R)\\[5pt]
        0                    &  j_\ell(k_d R) \eta_d     & h_\ell(k_d R) \eta_d        & j_\ell(k_m R)
\end{array}
\right]
\label{eqn:imdm}
\end{equation}

\noindent are modes of a MDM of radius $S=R+L$ with infinitely thick outer metal shell (labelled \emph{infinite} MDM.)\cite{Hasegawa2006Enhanced-surface} In our notation $j_\ell(x)$ ($h_\ell(x)$) is the spherical Bessel (Hankel) function of the first kind of integer order $\ell$, $\psi_\ell(x)=x j_\ell(x)$, $\xi_\ell(x)=x h_\ell(x)$ and the prime denotes differentiation with respect to the argument. The coupling terms $\left|\V{A}\right|$ and $\left|\V{B}\right|$ are obtained by replacing $j_\ell(k_m W)$ with $h_\ell(k_m W)$ in $\V{u}$ and $h_\ell(k_m S)$ with $j_\ell(k_m S)$ in $\V{V}$, respectively. For transverse electric (TE) modes $\eta_m =\eta_d =1$, while for transverse magnetic (TM) modes $\eta_m =\epsilon_m/\epsilon_0$ and $\eta_d =\epsilon_d/\epsilon_m$. The angular frequency of the incident EM field is $\omega$, such that $k_{0,m,d}=\sqrt{\epsilon_{0,m,d}}\ \omega/c$, with $c$ the speed of light in vacuum.

From Eq.~(\ref{eqn:mdm}) we immediately see that the resonances of the MDM system may be expressed in terms of coupled modes of the simpler resonators it is comprised of, as anticipated.\cite{Halas3} It is important to note here that these eigenfrequencies are in general complex, due to the radiative nature of the solutions.\cite{Sernelius} In the limit $x\gg\ell^{2}$ the mode coupling may be expressed using the asymptotic expansions of the Hankel function, $h_\ell(x) \sim (-i)^{\ell+1}e^{i x}/x$. Keeping $R$ and $L$ constant shows that the plasmon coupling between the inner sphere and outer shell decays exponentially with metal shell thickness $T$:

\begin{equation}
|\V{A}|\sim \left[ (-1)^\ell \left( \frac{\sqrt{\epsilon_m} - 1}{k_0W}\right) e^{i ( k_0  + k_m' )W + \pi/2 } \right] e^{-k_m'' W} 
    \label{eqn:sscouple}            
\end{equation}
where $k_m \equiv k_m' + i k_m''$. 

Based on several published algorithms, we have developed an efficient and stable method for calculating the EM scattering amplitudes and their corresponding coefficients for a large sphere with arbitrary number of alternating metal-dielectric layers.\cite{Hasegawa2006Enhanced-surface} We use this to investigate the EM energy distributions in the MDM sphere, as well as calculate its scattering cross section.

Using the procedure described above we obtain the eigenfrequencies of the three following systems: a metal sphere of radius $W=\nm{582}$, an infinite MDM with metal core radius $R=\nm{500}$ and dielectric shell of thickness $L=\nm{54}$, and a MDM system, with same core and dielectric shell dimensions as the infinite MDM and an finite outer metal shell of thickness $T=\nm{28}$. The thickness of the dielectric shell follows from the condition for a flat dispersion band,\cite{Hasegawa2006Enhanced-surface} $L \approx \pi c/(2\omega_{sp}\sqrt{\epsilon_d})$, where $\omega_{sp}$, the surface plasmon frequency is given by the resonance condition $\epsilon_m (\omega_{sp})+ \epsilon_d = 0$. In particular, we chose a silver--like Drude metal\cite{Hasegawa2006Enhanced-surface} whose dispersion is given by $\epsilon_m(\omega) = \epsilon_b - \omega_p^2(\omega^2 + i\Gamma \omega)^{-1}$. Here $\epsilon_b = 5.1$ describes the contribution of interband transitions, and $\hbar \omega_p = \unit[9.1]{eV}$ with $\omega_p$ the bulk plasma frequency. The free electron relaxation rate is given by $\Gamma$, where $\hbar \Gamma =\unit[0.021]{eV}$. The dielectric shell is chosen to be titania,\cite{Mohamed2003Properties-of-T} which is well approximated by the non-dispersive value of $\epsilon_d = 5.76$. Both the metal sphere and MDM are embedded in vacuum such that $\epsilon_0 = 1$.

\begin{figure}[t]
\begin{center}\includegraphics[width=\columnwidth]{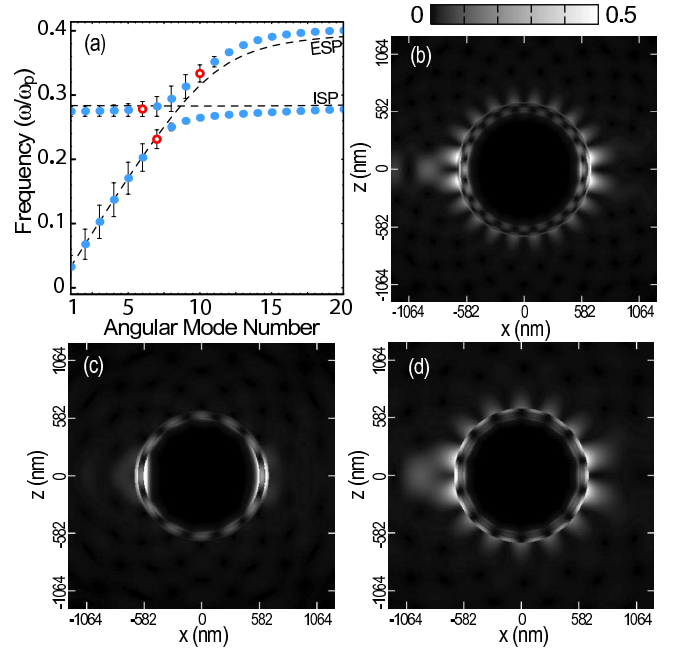}
\caption{(color online) (a) Angular mode dispersion of MDM microsphere. (b)-(d) Near-field energy density plots for the modes indicated by red circles in (a), at (b) $\omega/\omega_{p}=0.336$, (c) $\omega/\omega_{p}=0.294$ and (d) $\omega/\omega_{p}=0.233$.}
\label{fig:modes}
\end{center}
\end{figure}

In Fig.~\ref{fig:modes}(a) we plot the eigenfrequencies for the TM modes of the three geometries described above. The roots of Eq.~(\ref{eqn:mdm}) are labelled by a band index $n\geq 0$. The dashed line labelled ESP denotes the first band ($n=0$) of eigenfrequencies of the solid metal sphere obtained from Eq.~(\ref{eqn:ss}). The horizontal dashed line labelled ISP denotes the second band modes of the infinite MDM (Eq.~(\ref{eqn:imdm})). This band is obtained at the expected frequency\cite{Hasegawa2006Enhanced-surface} $\omega_{sp}/\omega_{p}=(\epsilon_b + \epsilon_d)^{-1/2}\approx0.3$. The two bands of solutions to Eq.~(\ref{eqn:mdm}) describing the MDM microsphere are also plotted, denoted by the blue circles. These bands correspond to $n=1$ and $n=2$. The $n=0$ band of this MDM (not plotted here) is at significantly lower frequencies, and hence does not couple to the higher order solutions. The width of each resonance is given by a vertical bar, equal in magnitude to twice the imaginary part of the eigenvalue, while the central frequency marked by the data points denotes its real part. The TE modes are all of frequencies greater than $\omega/\omega_{p}=0.4$ and are therefore not plotted here. A well-resolved avoided crossing is observed at $\omega/\omega_{p}=0.284$, a result of coupling between the solid metal sphere and the infinite MDM. 

Figures~\ref{fig:modes}(b)-(d) show grayscale plots of the near--field energy densities corresponding to three different frequencies in Fig.~\ref{fig:modes}(a). In Figs.~\ref{fig:modes}(b) and~\ref{fig:modes}(d) we observe that the field energies of the low and high frequency modes are concentrated at the outer metal shell, forming an \emph{external surface plasmon} (ESP) branch. For an intermediate frequency value, close to $\omega/\omega_{p}=0.284$ Fig.~\ref{fig:modes}(c) shows the field energy is concentrated at the interior shells' interfaces, thus belonging predominantly to an \emph{inner surface plasmon} (ISP) branch. These ESP and ISP branches coincide with the uncoupled solutions to the solid metal sphere and infinite MDM, respectively, accurately depicting the EM energy distributions in these systems. As we show below, suppression of the ESP in favor of excitation of the ISP results in a dramatic reduction of the MDM forward--scattering cross section.

\begin{figure}[t]
\begin{center}\includegraphics[width=\columnwidth]{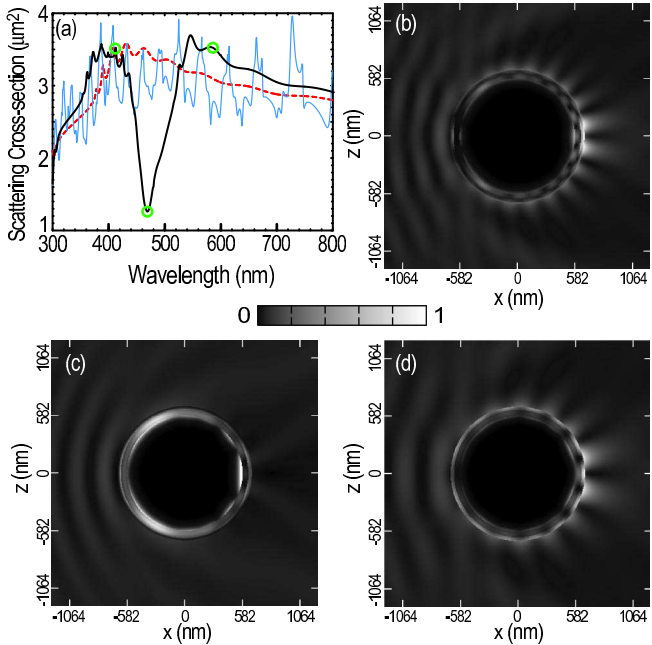}
\caption{(color online) (a) Scattering cross section of silver sphere (red, dashed), DMD (blue, thin line) and MDM (black, heavy line). (b)-(d) Near-field scattered-energy densities for (b) $\lambda=\nm{407}$, (c)$\lambda=\nm{463}$ and (d) $\lambda=\nm{586}$. All fields impinge from the left.}
\label{fig:xsec}
\end{center}
\end{figure}

In Fig.~\ref{fig:xsec}(a) we plot the scattering cross section, $C_{\rm sca}$ for the MDM described above. For comparison we also plot $C_{\rm sca}$ of the solid metal sphere discussed previously, as well as that of a DMD. The latter consists of a dielectric sphere surrounded by a metal shell, embedded in a dielectric.\cite{Halas1,Hasegawa2006Enhanced-surface} (It is not possible to compute $C_{\rm sca}$ for the infinite MDM since it does not support outward propagating solutions.) A large dip in $C_{\rm sca}$ is observed at a wavelength of $\lambda=\nm{463}$, corresponding to strong suppression of the forward--scattered fields. 

Energy densities of the scattered fields are shown in in Figs.~\ref{fig:xsec}(b)-(d), corresponding to circled wavelength values in Fig.~\ref{fig:xsec}(a). As expected, in Fig.~\ref{fig:xsec}(c) we observe strong suppression of the scattered field at $\lambda=\nm{463}$ ($\omega/\omega_{p}=0.294$) and the ISP nature of the fields is apparent. This suppression is reminiscent of coupled-resonator electromagnetically induced transparency (EIT).\cite{Boyd} Here too we see a transparency associated with large group delay, albeit in the MDM it is \emph{azimuthal}. Excitation of the ISP also suggests that EIT-like dynamical damping is responsible for the transparency. At higher and lower energies, where ESP excitations prevail the forward-scattered fields are significant, as seen in Figs.~\ref{fig:xsec}(b) and ~\ref{fig:xsec}(d). The effective scattering cross section at $\lambda=\nm{463}$ is equivalent to that of a silver sphere \nm{660} in diameter. Thus, adding one dielectric-metal nanoshell sequence onto a \um{1} diameter silver sphere reduces its EM footprint to that of a significantly smaller particle.

\begin{figure}[t]
   \centering
   \includegraphics[width=\columnwidth]{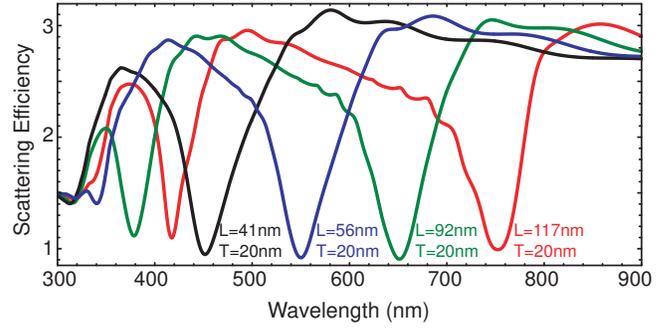}
   \caption{(color online) Tuning the scattering transparency window with dielectric shell width, $L$. The MDM consists of a silver core, $R=\nm{500}$, a titania shell of variable width $L$, and a silver shell of width $T=\nm{20}$.}
   \label{fig:tune}
\end{figure}

We further find that the scattering transparency may be tuned by adjusting the thickness of the dielectric shell. This is illustrated in Fig.~\ref{fig:tune}, where we plot the scattering efficiency, $C_{sca}/\pi W^2$ for several values of $L$. We use experimentally tabulated values for silver~\cite{Hasegawa2006Enhanced-surface} and amorphous titania~\cite{Mohamed2003Properties-of-T} to model a $R=\nm{500}$ silver core with a metal shell of fixed thickness $T$. By varying the dielectric shell thickness, $L$ the transparency is tuned across the entire visible spectrum. Figure~\ref{fig:tune} also shows the second--order transparency window of each of these spheres, arising from mode coupling in higher energy bands.

We note that the significant quantity typically measured in the far field is the extinction cross-section, describing the total energy removed from an incident plane wave due to scattering \emph{and} absorption.\cite{Bohren1983Absorption} For the MDM discussed here we have verified that the absorption peak is always offset in frequency from the transparency window, as well as being negligible compared to $C_{\rm sca}$. Thus our calculated scattering minima also result in true extinction transparencies in the far field.

In summary, we have shown that coupling of interior and exterior surface plasmon modes in MDM microspheres leads to resonant level splitting. This results in strong suppression of the forward--scattering cross section, to values much smaller than the geometric cross section of the particle. This transparency is spectrally tunable via the dielectric shell parameters. Such metal-dielectric coatings may be utilized to significantly reduce the EM footprint of large, non-planar metallic objects.

This work was supported by National Science
Foundation grant DMR-02-39273 and U.S. Army Research
Office grant DAAD19-02-1-0286. M. Deutsch's
e-mail is miriamd@uoregon.edu.

\end{document}